\documentclass{my_ws}

\begin{document}

\title{CO and Near-Infrared Observations of High-Redshift
Submillimeter Galaxies}

\author{D. T. Frayer}

\address{California Institute of Technology\\Astronomy
Dept. 105-24\\Pasadena, CA 91125\\E-mail: dtf@astro.caltech.edu}

\address{UMass/INAOE Conference on\\Deep Millimeter Surveys:
Implications for Galaxy Formation and Evolution\\19-21 June 2000}


\maketitle

\abstracts{I discuss our ongoing Owens Valley Millimeter Array
observations and Keck near-infrared wavelength observations of the
high-redshift sub-mm population of galaxies.  These observations are
important for our understanding of the distant universe since the sub-mm
population accounts for a large fraction of the extragalactic background
at mm/sub-mm wavelengths and contributes significantly to the total
amount of star-formation and AGN activity at high redshift.  The CO data
suggest that the sub-mm galaxies are analogous to the gas-rich
ultraluminous systems found in the local universe.  Initial
near-infrared data show that many of the sub-mm galaxies are faint-red
sources which are undetected at ultraviolet/optical wavelengths.  These
results highlight the importance that future sensitive mm-wavelength
instruments, such as the LMT and ALMA, will have on our understanding of
the early evolution and formation of galaxies.  }

\section{Introduction}

The discovery of an ultraluminous population of high-redshift galaxies
with deep submillimeter surveys has revolutionized our understanding of
the distant universe.\cite{sma97,bar98,hug98,eal99,bla99a} The current data
show that the sub-mm population has a mixture of AGN and starburst
characteristics with properties which are roughly consistent with the
local population of ultraluminous ($L>10^{12}L_{\odot}$) infrared
galaxies (ULIGs).  The relative importance of AGN and starburst activity
in powering the high luminosities of the sub-mm population is still an
open question, but the growing consensus is that the majority of the
luminosity of the population is powered by star formation.\cite{bla99b}
The early CO and X-ray data on the sub-mm population support the
starburst nature of the population by showing the presence of sufficient
molecular gas to fuel the star-formation activity\cite{fra98,fra99} and the
lack of expected X-ray emission if mostly dominated by
AGN.\cite{fab00,hor00,sev00,bar00}

Although the redshift distribution of the sub-mm population is still
uncertain, the majority of the sub-mm galaxies are believed to be at
high redshifts
($z\mathrel{\hbox{\rlap{\hbox{\lower4pt\hbox{$\sim$}}}\hbox{$>$}}}2$)
based on their radio\cite{cy99,sma00} and near-infrared\cite{sma99}
data.  The early redshift distributions based on optical imaging and
spectroscopy suggested somewhat lower redshifts,
\cite{sma98,bar99,lil99} but several of the original candidate optical
counter-parts have turned out to be incorrect.  Despite their ultra-high
luminosities, many sub-mm galaxies are undetected at ultraviolet/optical
wavelengths due to extinction by dust.  For these highly obscured
galaxies, follow-up radio\cite{sma00} and/or mm interferometric
observations\cite{dow99,fra00,ber00,gea00} are required in order to
uncover the proper counter-part.

In order to understand the nature of the sub-mm population, we have been
carrying out multi-wavelength observations of individual systems in the
SCUBA Cluster Lens Survey.\cite{sma98} This survey represents sensitive
sub-mm mapping of seven massive, lensing clusters which uncovered 15
background sub-mm sources. The advantages of this sample are that the
amplification of the background sources allows for deeper source frame
observations and that lensing by cluster potentials does not suffer
from differential lensing.  We have concentrated our efforts on the nine
background galaxies detected at the highest signal--to--noise (Table~1).
Only three sources have spectroscopic redshifts, and the redshift lower
limits shown in Table~1 are based on their sub-mm/radio flux
ratios.\cite{sma00}

\begin{table}[t]
\caption{Brightest Sources in SCUBA Cluster Lens Survey}
\begin{center}
\footnotesize
\begin{tabular}{ccccl}
\hline
\hline
\ & & & & \\
Galaxy & S(850$\mu$m)\cite{bar99} & Redshift & $K-$mag & Notes \\
  &(mJy) & & & \\ 
\hline
\ & & & & \\
SMM\,J02399$-$0136 &25.4&2.808~\cite{fra98}&19.1~\cite{ivi98}&CO\cite{fra98}\\
SMM\,J00266+1708 &18.6& $>2.0$~\cite{sma00} & 22.5~\cite{fra00}&\\
SMM\,J09429+4658&17.2&$>3.9$~\cite{sma00}&19.4~\cite{sma99}&ERO-H5\cite{sma99}\\
SMM\,J14009+0252 &14.5&$>0.7$~\cite{sma00}& 21.0~\cite{ivi00}&J5~\cite{ivi00}\\
SMM\,J14011+0252 &12.3&2.565~\cite{fra99}& 17.8~\cite{ivi00}&CO\cite{fra99}\\
SMM\,J02399$-$0134&11.0&1.062~\cite{sou99}& 16.3~\cite{sou99}&CO\cite{kne00}\\
SMM\,J22471$-$0206 & 9.2&$>1.8$~\cite{sma00} &?                 &\\
SMM\,J02400$-$0134 & 7.6&$>2.4$~\cite{sma00}  &?                 &\\
SMM\,J04431+0210 & 7.2&$>1.6$~\cite{sma00} &19.1~\cite{sma99}&ERO-N4\cite{sma99} \\
\hline
\end{tabular}
\end{center}
\end{table}

%
\begin{figure}[t]
\includegraphics{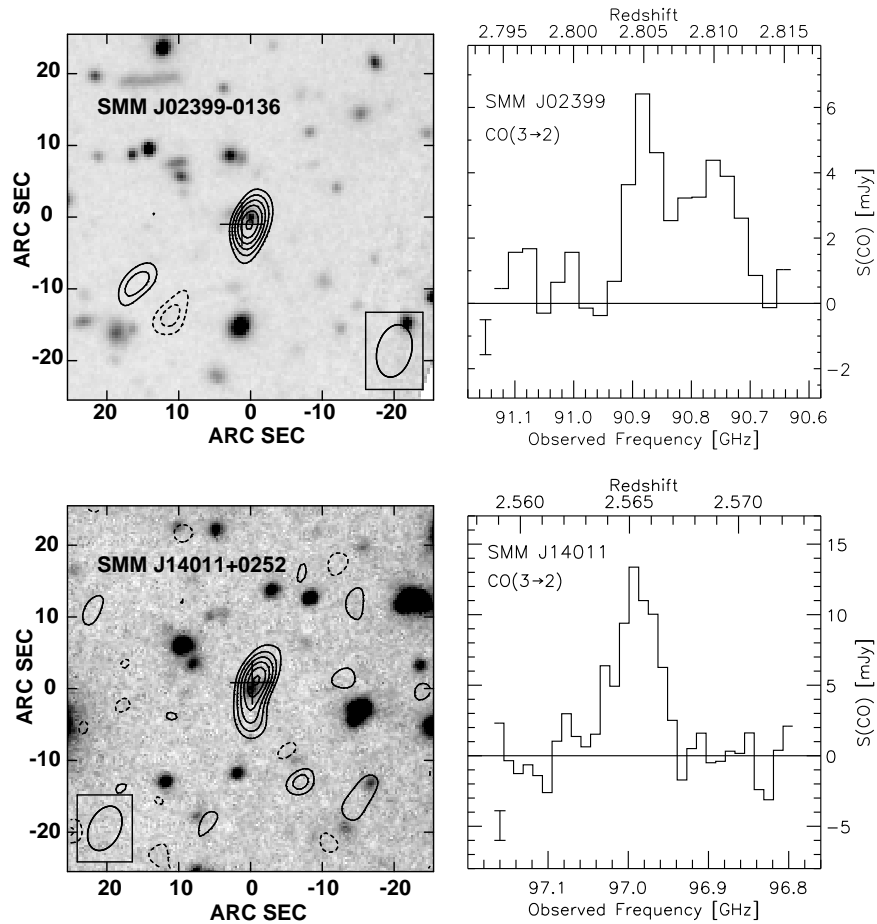}
\vspace*{5.0in}
\caption{OVRO CO(3-2) detections for SMM\,J02399$-$0136 (top) and
SMM\,J14011+0252 (bottom).  The grey-scale images at the left are
optical images while the contours represent the integrated CO maps for
the sub-mm galaxies. The crosses represent the positions of the SCUBA
detection.  The corresponding CO(3-2) spectra are shown at the right.
These data were originally published in Frayer et al. (1998,
1999).}
\end{figure}

\section{CO Results}

At OVRO we have conclusively detected CO emission from two sub-mm
systems, SMM\,J02399$-$0136 at $z=2.8$ (SMM\,J02399) and SMM\,J14011+0252
at $z=2.6$ (SMM\,J14011) (Fig. 1).  A third system SMM\,J02399$-$0134,
which is associated with a ring-galaxy containing a Seyfert nucleus at
$z=1.06$,\cite{sou99} has recently been detected in CO at the
PdB.\cite{kne00}  We have also tentatively confirmed the PdB detection
at OVRO.  To date, these three galaxies are the only sub-mm sources with
known redshifts, and it is promising that all three have already been
detected in CO.  The early CO results suggest that the sub-mm population
contains massive reservoirs of molecular gas and are among the most CO
luminous galaxies in the universe.

The strongest sub-mm source, SMM\,J02399, shows an AGN component in its
optical spectrum,\cite{ivi98} while SMM\,J14011 shows only evidence for
starburst activity at optical/NIR wavelengths.\cite{ivi00} Although the
optical characteristics of these two galaxies are vastly different,
their radio, sub-mm, and CO properties are fairly similar and are
consistent with a high level of star formation activity (SFRs of a
few$\times 10^{2}$\,M$_{\odot}$\,yr$^{-1}$ to more than
$10^{3}$\,M$_{\odot}$\,yr$^{-1}$, depending on the IMF and AGN
contamination).  After correcting for lensing, we derive CO luminosities
of 3--4$\times 10^{10}$\,K\,km\,s$^{-1}$\,pc$^2$
(H$_o=50$\,km\,s$^{-1}$\,Mpc$^{-1}$; $q_o=1/2$) in these two systems.
These CO luminosities correspond to molecular gas masses of about
$5\times 10^{10}$---$2\times10^{11}$\,M$_{\odot}$, depending on the
exact value of the CO to H$_2$ conversion factor.  Both SMM\,J02399 and
SMM\,J14011 appear to be associated with a merger event.  Given that
mergers of gas-rich galaxies at low-redshift result in massive
starbursts, we expect star-formation to be an important component for
powering the far-infrared luminosities in both of these systems.  In
fact, the large molecular gas masses of SMM\,J02399 and SMM\,J14011 are
sufficient to form the stars of an entire L$^{*}$ galaxy, which suggests
that the sub-mm population may represent the formative phase of massive
galaxies.

SMM\,J02399 is unresolved in CO, while SMM\,J14011 is extended over a
large spatial scale in its source frame
($\mathrel{\hbox{\rlap{\hbox{\lower4pt\hbox{$\sim$}}}\hbox{$>$}}}10$\,kpc).
Figure~1 contains low resolution OVRO data which showed tentative
evidence for extended CO emission in the north-south direction in
SMM\,J14011.  The extended morphology of the molecular gas in
SMM\,J14011 has been recently confirmed with higher-resolution data from
OVRO and BIMA.  These results may suggest that SMM\,J14011 is in an
early stage of its merger event, unlike the majority of ULIGs in the
local universe whose CO emission is mostly contained within the central
kpc.\cite{ds98,bs99} It is currently unknown what fraction of the sub-mm
sources are compact or are extended over large spatial scales as is
SMM\,J14011.  If the progenitors of the sub-mm systems are more gas rich
than those of local ULIGs, we could expect the sub-mm sources to have
larger gas fractions and to be more extended than their low-redshift
analogs.  A large sample of sub-mm sources need to be observed in CO
before statistical comparisons could be made between the CO properties
of local ULIGs and the high-redshift sub-mm galaxies.

%
\begin{figure}[t]
\includegraphics{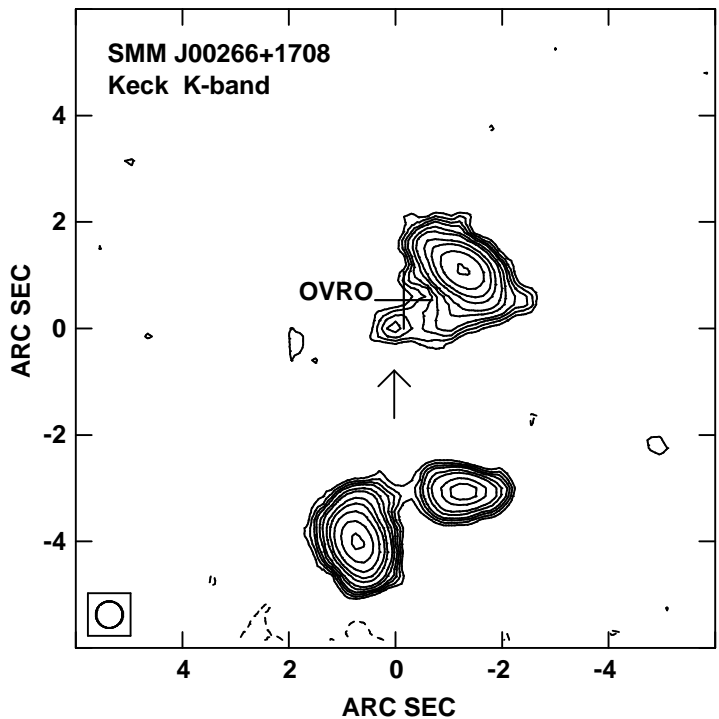} 
\vspace*{4.0in}
\caption{The Keck $K$-band ($2.2\mu$m) image of SMM\,J00266+1708.  The
near-infrared data was taken after determining the accurate position of
the sub-mm source with OVRO 1.3\,mm imaging.  The position of the
1.3\,mm source is shown by the cross labeled ``OVRO''.  The three bright
sources in the field are foreground galaxies previously observed at
optical wavelengths.  The arrow points to the new galaxy not detected in
the optical ($I>26$) thought to be the counter-part of SMM\,J00266+1708.
The rms of the image is 24.8 mag/sq-arcsec ($0.04\mu$Jy/beam), and the
contours are $1\sigma\times (-3,3,4,5,6,8,10,15,20,30,50,80)$.  The
seeing disk (beam) of the near-infrared data is shown in the lower left
($0.5\times0.5^{\prime\prime}$).  These data were originally published
in Frayer et al. (2000).}

\end{figure}

\section{Near-Infrared Results}

Many sub-mm galaxies are too obscured by dust to be detected at
ultraviolet/optical wavelengths.  At least four of the nine sources in
our sample which were undetected at optical wavelengths ($I>25$--26)
have faint near-infrared, $K$-band counter-parts.  Two of these are
bright enough in $K$-band ($K=19.1$, 19.6\,mag) to be classified as
extremely red objects (EROs).\cite{sma99} An additional faint ($K=21$)
galaxy was found associated with a relatively bright (0.5 mJy) radio
counter-part.\cite{ivi00} The fourth and faintest galaxy with a
near-infrared counter-part is SMM\,J00266+1708.

The sub-mm position of SMM\,J00266+1708 is located between three
optically visible galaxies.\cite{sma98} We imaged the field at OVRO at
1.3\,mm and determined its position to be offset from all three optical
sources.  Deep, follow-up near-infrared observations with NIRC on Keck
uncovered a new faint galaxy at $K=22.5$ located at the position of the
1.3\,mm source.\cite{fra00} Although SMM\,J00266+1708 is the second
brightest sub-mm source in the SCUBA Cluster Lens Survey, it is
currently the faintest known near-infrared counter-part of a sub-mm
galaxy discovered to date (Fig.~2).

Only two of the nine sources in the sample still require deep $K$-band
imaging and currently have uncertain counter-parts.  The galaxy
SMM\,J02400$-$0134 has no optically detected galaxies near the sub-mm
position.  For SMM\,J22471-0206 there are several optical galaxies which
could be the sub-mm counter-part, but given previous results it will be
interesting to test whether or not any new candidate galaxies are
uncovered with deep $K$-band imaging.  Depending on the results for
these last two unknown systems, the current data suggest that
approximately 40\%--70\% (4/9--6/9) of the sub-mm population as a whole
have faint near--infrared counter-parts which are undetected at optical
wavelengths.  Only 30\%--40\% (3/9--4/9) of the sample have optical
counterparts ($I<$26--27, correcting for source lensing).

The $K$-band magnitudes of the sub-mm counter-parts in the SCUBA Cluster
Lens Survey range over 6 magnitudes which reflects the wide diversity of
colors and redshifts for the population (Table~1).  The magnitudes
listed in Table~1 have not been corrected for lensing; unlensed sources
would be about a magnitude fainter on average.  Currently, only the
three brightest optical ($I$-band) sources have spectroscopic redshifts.
The other sources are much fainter and would require near-infrared
spectroscopy to obtain redshifts, which will be challenging even with
8m/10m class ground base telescopes.  Our early spectroscopic results
with NIRSPEC on Keck suggest that lines may be detectable in the
brighter sources
($K\mathrel{\hbox{\rlap{\hbox{\lower4pt\hbox{$\sim$}}}\hbox{$<$}}}20$),
while many of the fainter sub-mm sources
($K\mathrel{\hbox{\rlap{\hbox{\lower4pt\hbox{$\sim$}}}\hbox{$>$}}}22$)
may have to wait for the {\em Next Generation Space Telescope}.

\section{Conclusions}

Most sub-mm sources are too red and/or faint to be detected at optical
wavelengths.  There is very little overlap, if any, between the
ultraluminous sub-mm population and the less luminous,
optically-selected Lyman Break population of galaxies.  This highlights
the importance of radio, millimeter, and near-infrared observations of
the sub-mm population for our general understanding of the evolution and
formation of galaxies.

Potentially, we do not need to wait for future optical/near-infrared
space-based missions to obtain redshifts for the bulk of the sub-mm
population.  Redshifts could be determined directly from the CO lines
themselves with planned ground-based millimeter telescopes, such as the
LMT and ALMA.  Both ALMA and the LMT will have sufficient sensitivities
and broad-bandwidth spectrometer capabilities to make large CO redshift
surveys practical.\cite{bla00} The proposed 30~GHz spectrometer for the
LMT would be an excellent redshift machine for the sub-mm population of
galaxies.

The two best studied sub-mm galaxies (SMM\,J02399$-$0136 and
SMM\,J14011+0252) share many of the same properties of the local
population of ULIGs, such as high infrared luminosities, the association
with mergers, massive molecular gas reservoirs, comparable CO line
widths, and similar IR/radio and IR/CO luminosity ratios.  Future CO
observations of large samples of ultraluminous galaxies with ALMA and
the LMT will enable us to study the evolution of the molecular gas
properties as a function of redshift which will be crucial for our
understanding of the star-formation history of the universe.

\section*{Acknowledgments}

I thank the work of my collaborators Nick Scoville, Rob Ivison, Ian
Smail, Andrew Blain, Aaron Evans, Min Yun, and Jean-Paul Kneib. I
appreciate the efforts of the OVRO and Keck staff who have made these
observations a success.  I acknowledge support from NSF grant AST
9981546 made to the OVRO Millimeter Array which is operated by the
California Institute of Technology.  I thank the organizers at the
University of Massachusetts and INAOE for planning the conference and
providing support to attend the conference.

\end{document}